# Classification of large DNA methylation datasets for identifying cancer drivers

Fabrizio Celli[a], Fabio Cumbo[a,b], and Emanuel Weitschek[c,a]

[a]*Institute of Systems Analysis and Computer Science, National Research Council, Via dei Taurini 19, 00185 Rome, Italy*

[b]*Department of Engineering, Roma Tre University, Via della Vasca Navale 79, 00154 Rome, Italy*

[c]*Department of Engineering, Uninettuno International University. Corso Vittorio Emanuele II, 39 00186 Rome, Italy*

**Abstract**

DNA methylation is a well-studied genetic modification crucial to regulate the functioning of the genome. Its alterations play an important role in tumorigenesis and tumor-suppression. Thus, studying DNA methylation data may help biomarker discovery in cancer. Since public data on DNA methylation become abundant – and considering the high number of methylated sites (features) present in the genome – it is important to have a method for efficiently processing such large datasets. Relying on big data technologies, we propose BIGBIOCL an algorithm that can apply supervised classification methods to datasets with hundreds of thousands of features. It is designed for the extraction of alternative and equivalent classification models through iterative deletion of selected features.

We run experiments on DNA methylation datasets extracted from The Cancer Genome Atlas, focusing on three tumor types: breast, kidney, and thyroid carcinomas. We perform classifications extracting several methylated sites and their associated genes with accurate performance (accuracy>97%). Results suggest that BIGBIOCL can perform hundreds of classification iterations on hundreds of thousands of features in few hours. Moreover, we compare the performance of our method with other state-of-the-art classifiers and with a wide-spread DNA methylation analysis method based on network analysis. Finally, we are able to efficiently compute multiple alternative classification models and extract - from DNA-methylation large datasets - a set of candidate genes to be further investigated to determine their active role in cancer. BIGBIOCL, results of experiments, and a guide to carry on new experiments are freely available on GitHub at https://github.com/fcproj/BIGBIOCL.

# 1 Introduction

Tumor, or neoplasm, is a mass of tissue originated from an abnormal and uncontrolled division of eukaryotic cells. When tumoral cells invade and destroy surrounding tissues, the tumor is malignant and it is called cancer. According to the World Health Organization (http://www.who.int/mediacentre/factsheets/fs297/en/), nearly one six of death are caused by cancer. Since cancer is one of the leading causes of mortality, it is worth noting that research to fully understand its mechanisms and discover new ways to prevent and to treat this disease is fundamental to the human race. Transformation of healthy cells to tumoral ones is a complex process resulting from the interaction of genetic factors with external agents, like viruses, chemicals and physical mutagens. In this context, the importance of DNA methylation in carcinogenesis is widely recognized (Baylin, 2005; De Carvalho et al., 2012; Feinberg et al., 2006; Figueroa et al., 2008; Zhuang et al., 2002).
DNA Methylation is one of the most intensely studied genetic modification in mammals involving reversible covalent alterations of DNA nucleotides (Bird, 2002). In particular, the enzyme DNA methyltransferase catalyzes the conversion of the cytosine (typically in a CpG site) to 5-methylcytosine, by adding a methyl group ($CH_3$) to cytosine residues in the sequence. In normal cells, this conversion results in different interaction properties assuring the proper regulation of gene expression and of gene silencing (Baylin et al., 2001). In the haploid human genome there around 28 million of CpG sites in methylated or unmethylated state (Stevens et al., 2013). It is well-known that inactivation of tumor-suppressor genes may occur as a consequence of hyper-methylation within the gene regions and a large range of cancer-related genes can be silenced by DNA methylation in different types of tumors. Moreover, a global hypo-methylation, which induces genomic instability, also contributes to cell transformation (Kulis et al. 2010). Thus, methylation corresponds to inactivity, but inactivity of a repressive factor means stimulation. This means that studying DNA methylation data to identify drivers in cancer is challenging.
Another challenge is given by the reduction of the cost of data generation that, especially after the employment of Next-Generation Sequencing technologies (Weitschek et al., 2014), has made available an enormous amount of raw data. The availability of big datasets creates problems with the application of classical algorithms for data mining and analysis (Greene et al., 2014).

In this work, we focus on the adoption of big data technologies for the application of classification algorithms on large DNA methylation datasets. Even if there are many different definitions of big data, "Big data refers to datasets whose size is beyond the ability of typical database software tools to capture, store, manage and analyze" (McKinsey Global Institute, 2012). This definition does not focus on specific data size, but on the technology we adopt to manage those datasets.

We want to extract a set of genes that may play a role in a specific tumor by applying supervised learning methods to DNA methylation datasets with a large number of features (450 thousand CpG positions). We aim to compute many classification models containing genes by applying optimized supervised learning algorithms, like Decision Trees (Quinlan, 1993) and Random Forests (Breiman, 2001; Svetnik et al., 2003). We rely on Apache Spark MLlib (Meng et al., 2016), running in standalone or cluster mode, in order to cope with performance. In fact, the largeness of the input dataset does not allow to analyze and process it in an acceptable time with non-big data technologies.
A previous classification study on DNA methylation (Danielsson et al., 2015) proposed MethPed, a tool for the identification of pediatric tumors. Researchers built the classification model behind MethPed from DNA methylation datasets with 450 thousand of features. They firstly applied a large number of regression algorithms to select a subset of features with the highest predictive power; then, they adopted Random Forests to build the classification model. On the contrary, we want to apply classification algorithms to the entire dataset in order to obtain a large number of CpG sites and their associated genomic locations. Another study (Akalin et al., 2012) described methylKit, an R package for the analysis of DNA methylation data. This package adopts an unsupervised machine learning approach, working on unlabeled data. methylKit works in-memory and, even if it is multi-threaded, its execution is limited to a single machine. On our side, we want to perform supervised machine learning on a cluster of computational nodes, in order to be able to scale with the increasing dimension of input data.

The algorithm proposed in our study is inspired by CAMUR for being applied to large input datasets. CAMUR (Classifier with alternative and multiple rule-based models) is a classification method that iteratively computes a rule-based classification model, eliminates from the input dataset combinations of extracted features, and repeats the classification until a stopping condition is verified (Cestarelli et al., 2016). The result of a CAMUR computation is a set of classification models. CAMUR worked on RNA sequencing cancer datasets with around 20 thousand features. In this work, we design and develop BIGBIOCL, a multiple tree-based classifier, to analyze DNA methylation datasets with more than 450 thousand features (Pidsley R. et al., 2013). Our goal is to extract candidate methylated sites and their related genes in few hours.

# 2 Methods

In our experiments on the application of big data technologies to the classification of large DNA methylation datasets, we consider three types of cancer: the Breast Invasive Carcinoma (BRCA), the Thyroid Carcinoma (THCA), and the Kidney Renal Papillary Cell Carcinoma (KIRP). We develop BIGBIOCL in order to run an iterative classification algorithm in big data environments, to achieve efficient supervised learning, and to extract multiple classification models. Then, we test our algorithm both in a single-machine and in a Hadoop YARN cluster.

## 2.1 Datasets

The Cancer Genome Atlas (TCGA) is a project started in 2005 and maintained by the National Cancer Institute and National Human Genome Research Institute (Weinstein et al., 2013). The TCGA is a 2.5 petabytes public dataset widely used in scientific research. Searching "The Cancer Genome Atlas" on PubMed reveals more than 2,500 articles in the last 5 years. The TCGA dataset contains the genomic characterization of over 30 types of human cancer (Tomczak *et al*., 2015) from more than 11,000 patients. The dataset includes cancer genome profiles obtained from several NGS methods applied to patient tissues, like RNA sequencing, Array-based DNA methylation sequencing, microRNA sequencing, and many others (Hayden 2014, Weitschek et al., 2014, Shendure, J. et. al 2008) .

In our work, we focus on DNA methylation data. In particular, we consider profiles obtained using the Illumina Infinium Human DNA Methylation 450 platform (HumanMethylation450), which provides quantitative methylation measurement at CpG site level (Sandoval et al., 2011). HumanMethylation450 allows assessing the methylation status of more than 450 thousand CpG sites (Dedeurwaerder et al., 2014), producing large datasets to be analyzed and interpreted. Even if HumanMethylation450 datasets can be useful for large-scale DNA methylation profiling, they raise problems of efficient data processing.

Consequently, we decide to explore the adoption of big data technologies and infrastructures to enable the possibility of efficiently applying machine learning algorithms to such large datasets. We rely on the latest TCGA data release available at The Genomic Data Commons data sharing platform (https://gdc.nci.nih.gov/).

In our experiments, we use the *beta value* as an estimate of DNA methylation level. Beta value (Du P. et al., 2010) is defined as the ratio of the methylated allele intensity and the overall intensity (i.e. the sum of methylated and unmethylated allele intensities):

$$\beta_n = \frac{\max(Meth_n, 0)}{\max(Meth_n, 0) + \max(Unmeth_n, 0) + \varepsilon} \quad (1)$$

where $Meth_n$ is the n$^{th}$ methylated allele intensity, $Unmeth_n$ is the n$^{th}$ unmethylated allele intensity, and $\varepsilon$ is a constant offset used to regulate the beta value where both intensities are low. It is worth noting that beta value is a continue variable in the range [0, 1], where 0 means no methylation and 1 full methylation.

**Table 1.** Datasets used in this study

| Dataset | Number of Samples | Number of Features |
|---------|-------------------|--------------------|
| BRCA    | 897               | 485,512            |
| THCA    | 571               | 485,512            |
| KIRP    | 321               | 485,512            |

We focus on three DNA methylation datasets extracted from TCGA: BRCA, THCA, and KIRP (Table 1). For each dataset, we filter the input data matrix to cope with missing values and to exclude control cases (this is important to reduce the classification task to binary classification, having only tumoral and normal cases). The final data matrix (Table 2) has the following structure:

- Rows represent samples, i.e. the profile of a patient tissue. The first row is the header, so it contains column names.
- The first column contains ID of samples. The last column is the category, specifying if the sample is "tumoral" or "normal".
- All other columns represent CpG sites, and the corresponding cells contain the beta value for the CpG site. We use the Illumina 450k manifest to know where a CpG site is located and which gene corresponds to it. The manifest is available on Illumina website (https://support.illumina.com/array/array_kits/infinium_humanmethylation450_beadchip_kit/downloads.html).
- Missing values are encoded with the question mark.

**Table 2.** Structure of the DNA methylation data matrix extracted from TCGA

| Sample ID       | cg13869341 | … | cg00381604 | Class   |
|-----------------|------------|---|------------|---------|
| TCGA-A7-A0DC-11 | 0.971644   | … | 0.017485   | Tumoral |
| TCGA-BH-A0BV-11A| 0.925557   | … | ?          | Normal  |
| TCGA-BH-A0DZ-11A| 0.907020   | … | 0.019204   | Tumoral |

### 2.2 Supervised Learning

The goal of our study is to develop an iterative algorithm that can efficiently extract a set of genes from large DNA methylation cancer datasets. The first step is the application of a supervised learning method (Tan et al., 2005; Weitschek et al., 2014). This is possible because the datasets used in this study (Table 1) are labeled datasets, i.e., we know if each tissue belongs to the 'normal' or 'tumoral' category. Using a labeled dataset (or a part of it) as a training set, the supervised learning algorithm infers some hypothesis from the features and builds a classification model, which is simply a function that assigns a category to a sample. We perform tests with both Decision Trees (Quinlan, 1993) and Random Forests (Breiman, 2001; Svetnik et al., 2003). Then, we extract CpG sites (features) from the classification model and the corresponding genes. The list of genes extracted from a classification model is part of the output of our algorithm. In fact, as we explain in the next section, our algorithm runs many iterations, and the overall result is the union of the results of each iteration. It is important to highlight that we are not interested in the decision model to classify new data (even if this would be possible), but to extract a list of candidate genes that may play a role in cancer.

Decision Trees are used for recursive binary partitioning of the feature space. Starting from the root, which contains the entire training dataset, Decision Trees are built by splitting the dataset into distinct nodes, where a node defines the probability of a point to be of a certain category. The final prediction is the label of the final leaf node reached during the decision process. Decision Trees are smooth to understand and they allow validating the model with statistical tests (like entropy or information gain). Unfortunately, it is easy to create a tree that overfits the input data. In addition, since Decision Trees use a greedy algorithm, the optimal tree is sometimes not found.

Random Forests solve many problems of Decision Trees, especially when applied to very large datasets. Random Forests run many Decision Trees in parallel and they fit well with big data technologies and map-reduce algorithms, since data can be split on different machines. There are two points of randomness that reduce the possibility of overfitting and over generalization. First of all, each tree is created from a random selection of N data points from the training set. Then, during the decision process of a specific tree, there is a random selection of M features from the global set of features. For all those reasons, while both Decision Trees and Random Forests are explored, the final implementation of BIGBIOCL is based on Random Forests.

### 2.3 BIGBIOCL: a multiple tree-based classifier for big biological data

CAMUR (Cestarelli et al., 2016) is a supervised method that can extract alternative and equivalent classification models from a labeled dataset (Weitschek, 2016). CAMUR adopts an iterative feature elimination technique: it uses the supervised RIPPER algorithm (Cohen, 1995) to compute a rule-based classification model, iteratively eliminates combinations of features that appear in the model from the input dataset, and performs again the classification until a stopping condition is verified. Once a feature is eliminated from the dataset, it can be reinserted in the next iteration (*loose* execution mode) or discarded forever (*strict* execution mode). CAMUR has been successfully applied to RNA-sequencing data (Cestarelli et al., 2016) extracted from TCGA, and evaluated on Gene Expression Omnibus (GEO) datasets. Datasets used in CAMUR tasks contained at most 30 thousand of features and a thousand of samples. When trying to apply CAMUR to DNA methylation datasets, which contain hundreds of thousands of features, the algorithm suffers of memory and execution time problems.

In this work we propose BIGBIOCL, a JAVA command-line software that is inspired by CAMUR to enable the efficient management and classification of large datasets. BIGBIOCL adopts big data solutions and introduces many innovations to CAMUR:

- BIGBIOCL is based on MLlib, the Apache Spark's scalable machine learning library. The adoption of Apache Spark allows executing the algorithm on Hadoop YARN (Vavilapalli et al. 2013) cluster, with the possibility to parallelize the machine learning task on several machines.
- Even if both Decision Trees and Random Forests have been tested, the final implementation of BIGBIOCL is based on Random Forests. One of the reasons is that, Random Forests naturally fit with parallel computation, since each node of a cluster can compute a different tree of the forest and send the result back to a master node.
- BIGBIOCL, following the CAMUR method, iteratively computes a Random Forest model. After each iteration, BIGBIOCL permanently removes all features that appear in the computed model from the input dataset, and not only combinations of them. This approach is similar to the CAMUR loose execution mode, but removing all extracted features makes the entire process lighter since there is no more the need to compute the power set at each iteration. Obviously, having hundreds of thousands of features guarantees that a relevant number of alternative classification models are still extracted, as we show in the next section.

BIGBIOCL iterative procedure stops when the reliability of the classification model is below a given threshold, or when a maximum number of iterations has been reached. Both stopping conditions must be specified by the user as command-line parameters. We use the F-measure to evaluate the accuracy of classification models. The F-measure is defined as the weighted harmonic mean of precision (P) and recall (R). We decide to equally weight precision and recall, obtaining the formula:

$$F - \text{measure} = \frac{2PR}{P+R} \qquad (2)$$

It is worth noting that F-measure is high when both precision and recall are high. Precision and recall are defined in terms of true positive TP (the number of samples that are assigned to a category and that belong to that category), false positives FP (the number of samples not belonging to a category but assigned to that category), and false negatives FN (the number of samples belonging to a category but not assigned to that category):

$$P = \frac{TP}{TP+FP} \; ; \; R = \frac{TP}{TP+FN} \qquad (3)$$

When the iterative algorithm stops, the software collects the list of features that appear in all computed classification models. Since features are CpG sites that are located in different genomic regions, we use a mapping file for discovering the gene where a CpG site is located (see section 2.1 for further details). The software can therefore derive a list of candidate genes as final output of the computation, associating them to the tumor under study. Extracted genes can then be explored and evaluated by biologists to investigate their role in cancer. Obviously, BIGBIOCL can be applied also to different datasets. In fact, it is not limited to DNA methylation data, but it works on any input dataset having the structure illustrated in Table 2.

## 3 Results

In this section, we discuss the path that led to the Random Forests implementation of BIGBIOCL, providing statistics about experiments and a discussion about results. All our experiments refer to the datasets listed in Table 1.

First of all, we tried to use CAMUR in *strict* mode to extract candidate genes from the BRCA dataset. As we have previously noted, CAMUR works properly with TGCA RNA-sequencing data, where the number of features is around 30 thousand. The BRCA dataset - stored in a 6.5GB text file - includes more than 450 thousand of features and CAMUR cannot manage such amount of data. The experiment was executed using the workstation described in Table 3, allocating 22GB of RAM and 7 cores to the Java Virtual Machine (JVM). After 16 minutes, CAMUR ran out of memory.

**Table 3.** Workstation used for experiments

| Parameter | Value |
|---|---|
| Architecture | x86 |
| CPU | Intel(R) Core(TM) i7-4790 CPU @ 3.60GHz |
| Number of CPUs | 8 |
| RAM | 24GB |
| OS | CentOS Linux release 7.3.1611 |
| Java Version | Oracle jdk1.8.0_131 |

Afterwards, we executed several experiments, relying on Apache Spark MLlib:

(1) Single iteration of Decision Trees. We ran Decision Trees in Spark local mode, in order to evaluate results and performance.
(2) Single iteration of Random Forests. We ran Random Forests in Spark local mode, in order to compare results and performance with Decision Tree experiments.
(3) Execution of Linear Support Vector Machines (SVMs) and Naïve Bayes. We ran SVMs and Naïve Bayes in Spark local mode to compare the accuracy of Random Forests results with other classification methods.
(4) BIGBIOCL: this is the Random Forest iterative algorithm (with feature deletion) implemented with big data technologies. The algorithm was tested both in Spark local mode and on Apache Hadoop YARN multi-node cluster.

Apache Spark local mode is a non-distributed single-JVM configuration that allows Spark to run all its execution components (i.e. driver, executor, scheduler, and master) in the same JVM. In local mode, the default parallelism is the number of threads specified as command line parameter. Table 4 and Table 5 show the configuration and results of experiments with a single iteration of Decision Trees in the same workstation used for testing CAMUR. In all our experiments we used 70% of randomly sampled input data to build the model (training set), and 30% of data for the evaluation (test set). Results show that BIGBIOCL can manage large datasets with hundreds of thousands of features.

**Table 4.** Configuration of Decision Tree experiments - Spark local mode (dataset: BRCA)

| ID | Memory | Threads | Max Depth | Max Bins | Impurity |
|---|---|---|---|---|---|
| 1 | 5 GB | 4 | 5 | 16 | Gini |
| 2 | 5 GB | 4 | 5 | 32 | Gini |
| 3 | 12 GB | 7 | 5 | 32 | Gini |
| 4 | 12 GB | 7 | 10 | 32 | Gini |
| 5 | 12 GB | 7 | 5 | 8 | Gini |
| 6 | 18 GB | 7 | 5 | 128 | Gini |

Experiments with Decision Trees demonstrate that we were able to classify large datasets, even using only 5 GB of memory. The execution time decreases drastically if the parameter *max bins* is reduced. Even if execution time seems to be acceptable (the algorithm terminates at most in one hour), some observations led us to test (and then adopt) Random Forests:

- We could extract only few features from each execution of the algorithm. We are interested in identifying a set of candidate genes for a specific type of cancer, thus having more features would be preferable.
- Decision Trees offer few possibilities of parallelization. This is important especially in the context of multiple iterations, where parallelization can reduce the overall execution time. On the other hand, Random Forests allow splitting the data on many machines, reducing the execution time of each iteration.

**Table 5.** Results of Decision Tree experiments described in Table 4

| ID | Build Time | Evaluation time | F-Measure | #Features |
|----|------------|-----------------|-----------|-----------|
| 1 | 37.7 min | 17.5 min | 98,51% | 2 |
| 2 | OOM | - | - | - |
| 3 | 66.23 min | 1.96 min | 98.76% | 4 |
| 4 | 67.96 min | 1.92 min | 99.20% | 4 |
| 5 | 9.6 min | 1.92 min | 98.03% | 3 |
| 6 | OOM | - | - | - |

This is Table shows the execution time and results of a single iteration of Decision Trees. The configuration adopted for each experiment is provided in Table 4. "ID" is the unique identifier for an experiment. "Build Time" is the time needed to build the classification model, while "Evaluation time" is the time for the evaluation of the model on test data (30% of input data). The accuracy of the model is given by the F-measure. The column "#Features" represents the number of features that appear in the classification model, i.e. the CpG loci that can be extracted. "OOM" means that the experiment ran out of memory.

Table 6 and Table 7 show results of some experiments with Random Forests. Overall, a single execution of Random Forests performs definitely better than a single execution of Decision Trees. Even if experiment 7 produced a result in more than one hour and a half, experiment 8 shows that increasing the memory from 5 GB to 12 GB dramatically improves the execution time. To build the model, experiment 8 required 38.5% of the time of the equivalent experiment with Decision Trees (ID=3). In addition, Random Forests produce more features, which is important to identify more genes that may play a role in cancer.

**Table 6.** Configuration of Random Forest experiments - Spark local mode (dataset: BRCA)

| ID | Memory | Threads | Max Depth | Max Bins | #Trees | Impurity |
|----|--------|---------|-----------|----------|--------|----------|
| 7 | 5 GB | 7 | 5 | 16 | 5 | Gini |
| 8 | 12 GB | 7 | 5 | 16 | 5 | Gini |
| 9 | 12 GB | 7 | 5 | 16 | 10 | Gini |

**Table 7.** Results of Random Forest experiments described in Table 6

| ID | Build Time | Evaluation time | F-Measure | #Features |
|----|------------|-----------------|-----------|-----------|
| 7 | 1 h 35 min | 20.37 min | 98,92% | 33 |
| 8 | 25.53 min | 1.73 min | 98.47% | 40 |
| 9 | 28.87 min | 1.97 min | 98.83% | 77 |

Additional experiments with other methods for large-scale classification tasks, i.e. Support Vector Machines (SVMs) and Naïve Bayes, justify the adoption of Random Forests in the final implementation of BIGBIOCL. Comparing Tables 7 and 9, we observe that experiments with SVMs show greater execution times than experiments with Random Forests. Even varying the amount of RAM (from 5GB to 18GB), execution time of SVMs does not change. In addition, while F-Measures of Tables 7 and 9 are comparable, SVMs do not provide a human interpretable model that we can use to create a list of candidate genes. We have also performed experiments with multinomial Naïve Bayes, but F-Measures were much lower and we could not rely on a human interpretable model to extract relevant features.

**Table 8.** Configuration of SVM experiments - Spark local mode (dataset: BRCA)

| ID | Memory | Threads | Regularization method | Regularization parameter | #Iterations |
|----|--------|---------|----------------------|-------------------------|-------------|
| SVM1 | 12 GB | 7 | L2 | 1.0 | 100 |
| SVM2 | 12 GB | 7 | L2 | 1.0 | 200 |
| SVM3 | 12 GB | 7 | L1 | 0.1 | 100 |
| SVM4 | 12 GB | 7 | L1 | 0.1 | 200 |

**Table 9.** Results of SVM experiments described in Table 8

| ID | Execution Time | F-Measure |
|----|----------------|-----------|
| SVM1 | 2 h 03 min | 98.95% |
| SVM2 | 3 h 32 min | 98.74% |
| SVM3 | 1 h 40 min | 95.46% |
| SVM4 | 1 h 19 min | 99.16% |

For comparing the performance of our algorithm with a sequential implementation of Decision trees and Random Forest classifiers, we decided to run the classification analyses by adopting the Weka software package (Hall et al., 2009). The amount of memory we had to allocate was 24GB in order to permit the execution of the sequential algorithms. The running time of the sequential Random Forest on BRCA was 15.5 minutes (model building and evaluation) setting the max bins to 2, the number of trees to 20, and the max depth to 5 obtaining an F-Measure value of 98.33%. Conversely, the Random Forest Apache Spark single node implementation with the same settings took 8 minutes with an F-Measure value of 99.81%.

Moreover, when testing the sequential implementation of the Decision Tree results are even more noteworthy. A run of the sequential implementation with the same settings of experiment 5 in Table 4 did not compute a solution even after 20 days of computation, while the Apache Spark implementation terminated just in 10.5 minutes. Finally, it is worth noting that both Spark implementations need less memory (12GB and 18 GB) to perform the classification analyses.

BIGBIOCL was tested both running Apache Spark in local mode and on Apache Hadoop YARN Cluster. Experiments with Hadoop Cluster were performed using PICO (http://www.hpc.cineca.it/hardware/pico), the latest Cineca's Italian Supercomputing infrastructure for big data. PICO allows allocating computational nodes and memory on demand when running Hadoop jobs (Table 10). Experiments with Spark in local mode were conducted in the workstation described in Table 3. In both cases, we used Apache Spark 2.1.1. BIGBIOCL implements the following iterative algorithm:

- At each iteration, Apache MLlib Random Forests model is computed on the working dataset S. On a cluster, trees of the Random Forests can be computed in parallel on different nodes.
- At iteration 0, S is equals to the input dataset.
- After each iteration, the set of features F that appear in the computed model is removed from S. Thus, next iteration runs on the dataset $\{S - F\}$.
- Once eliminated, features are never reintegrated in the working dataset S.
- The algorithm terminates when F-measure on test data is below a threshold MF (parameter provided by the user) or the number of iterations is bigger than a threshold MI (in the rest of this article we consider $MI = 1000$).

Tables 11 and 12 show results of experiments running Spark in local mode. The input dataset is BRCA. As we can see, setting the F-measure threshold to 99%, BIGBIOCL ran 2 iterations in around one hour, extracting 224 candidate genes. Relaxing that constraint, we had more iterations and more candidate genes. When the F-measure threshold was set to 97%, BIGBIOCL executed 96 iterations, computing 5072 genes in less than 2 days. Experiments on Hadoop YARN Cluster are summarized in Tables 13 and 14. They were useful to evaluate how performance improves with parallelization on multiple computational nodes. Results are attractive. Experiment 16 (on Hadoop) corresponds to Experiment 12 (Spark local mode) and its execution time was 22% of Experiment 12. It is also interesting to note (Experiments 16 and 17) that increasing the number of working nodes of the cluster (so also the total number of CPUs) we got more iterations and more genes.

We wish to highlight that all the extracted genes related to each tumor are available at supplementary data S1. Additionally, a comprehensive description of the experimentation is provided in the wiki of BIGBIOCL on GitHub.

Furthermore, if we compare Experiment 18 (on Hadoop) with Experiment 13 (Spark local mode), we can notice again how the execution on a cluster outperforms the Spark local mode, both in terms of execution time and of number of features extracted. On average, running BIGBIOCL in Spark local mode requires around 1500 seconds to generate the classification model at each iteration, while using 3 PICO's nodes on Hadoop YARN cluster the average time to build a classification model is 330 seconds.

**Table 10.** PICO's hardware, used for experiments with Hadoop Cluster

| Parameter | Value |
|---|---|
| Total Nodes | 66 |
| CPU | Intel Xeon E5 2670 v2 @2.5Ghz |
| Cores per node | 20 |
| RAM per node | 128 GB |

Tables 15 and 16 show results of some experiments with THCA and KIRP datasets. Experiments refer to the execution of BIGBIOCL in Spark local mode, using the workstation described in Table 3. On average, the time to build a classification model for the KIRP dataset is 340 seconds, while for THCA this number increases to 945 seconds. This result is quite obvious, since THCA contains 571 samples, while KIRP only 321. What is interesting to note is that on THCA the algorithm stops after 7 iterations, while on KIRP after 34 iterations, even if the KIRP dataset contains less samples. This depends on the different distribution of beta values in the two datasets.

For estimating the execution time of a sequential implementation of our algorithm, we can consider experiment number 17 (Tables 13 and 14) whose execution time was 13 h 30 min. If we run the same number of iterations (i.e., 116) with the sequential implementation of Random Forest of the Weka software package, the execution time will be at least of 30 h (not taking into account potential overhead).

**Table 11.** Configuration of BIGBIOCL experiments - Spark local mode (dataset: BRCA)

| ID | Memory | Threads | Max Depth | Max Bins | #Trees | Stopping Condition |
|---|---|---|---|---|---|---|
| 10 | 18 GB | 7 | 5 | 16 | 5 | F-measure < 98% |
| 11 | 18 GB | 7 | 5 | 16 | 10 | F-measure < 98% |
| 12 | 18 GB | 7 | 5 | 16 | 10 | F-measure < 97% |
| 13 | 18 GB | 7 | 5 | 16 | 20 | F-measure < 99% |

**Table 12.** Results of BIGBIOCL experiments described in Table 11

| ID | Overall Time | #Iterations | #Features | #Distinct Genes |
|---|---|---|---|---|
| 10 | 3 h 33 min | 8 | 331 | 230 |
| 11 | 13 h 16 min | 26 | 2345 | 1460 |
| 12 | 46 h 34 min | 96 | 9780 | 5072 |
| 13 | 1 h 2 min | 2 | 329 | 224 |

"Overall time" is the time to execute all iterations. For each iteration, execution time includes the time to build the model, the time to evaluate the model on test data, and the time to evaluate the model on training data.

**Table 13.** Configuration of BIGBIOCL experiments - Hadoop YARN Cluster (dataset: BRCA)

| ID | #Nodes | Mem per node | CPU per node | Max depth | Max bins | #Trees | Stopping Condition |
|---|---|---|---|---|---|---|---|
| 14 | 2 | 96 GB | 20 | 5 | 16 | 5 | F-measure < 98% |
| 15 | 2 | 96 GB | 20 | 5 | 16 | 10 | F-measure < 98% |
| 16 | 2 | 96 GB | 20 | 5 | 16 | 10 | F-measure < 97% |
| 17 | 3 | 96 GB | 20 | 5 | 16 | 20 | F-measure < 97% |
| 18 | 3 | 96 GB | 20 | 5 | 16 | 20 | F-measure < 99% |

"#Nodes" is the number of PICO's nodes allocated to the execution of the experiment. For each working node, we specified an amount of memory ("Mem per node") and the number of CPU ("CPU per node").

**Table 14.** Results of BIGBIOCL experiments described in Table 13

| ID | Overall Time | #Iterations | #Features | #Distinct Genes |
|---|---|---|---|---|
| 14 | 28.58 min | 4 | 165 | 123 |
| 15 | 1 h 56 min | 16 | 1352 | 907 |
| 16 | 10 h 36 min | 88 | 8722 | 4607 |
| 17 | 13 h 30 min | 116 | 24984 | 9539 |
| 18 | 22.15 min | 3 | 507 | 352 |

**Table 15.** Configuration of BIGBIOCL experiments - Spark local mode (datasets: THCA and KIRP)

| ID | Memory | Threads | Max depth | Max bins | #Trees | Stopping Condition |
|---|---|---|---|---|---|---|
| THCA 19 | 18 GB | 7 | 5 | 16 | 5 | F-measure < 97% |
| KIRP 20 | 18 GB | 7 | 5 | 16 | 5 | F-measure < 97% |

**Table 16.** Results of BIGBIOCL experiments described in Table 15

| ID | Overall Time | #Iterations | #Features | #Distinct Genes |
|---|---|---|---|---|
| THCA 19 | 2 h 13 min | 7 | 541 | 398 |
| KIRP 20 | 4 h 21 min | 34 | 1215 | 852 |

In order to compare our results with a wide-spread DNA methylation analysis method, we followed the procedure described in (Bartlett, *et al.* 2014). We have computed all the pairwise Pearson correlation coefficients (PPCC) between all CpG islands in the three examined datasets (BRCA, KIRP, and THCA). The aim of this operation was to construct a correlation network for each tumor differentiating normal and tumoral tissues.

To achieve this goal, a cleaning of the dataset was required. In particular, for each tumor, we have replaced the unavailable measurements in our datasets with the mean value computed on the known beta values. Additionally, because of the nature of this analysis, we have fixed a threshold at 0.9 on the correlation measure to identify the strong correlated CpG islands only. This means that, if the correlation between the island X and the island Y is greater than 0.9 (in module), an edge between X and Y will be inferred.

It is worth noting that this kind of analysis was extremely time consuming due to the dimension of our datasets and due to the non-parallel implementation of the method described in (Bartlett, *et al.* 2014), as shown in Table 17.

Due to the small dimension of the inferred networks (Figure 1, 2, 3), any analytical method from network theory is useful, except in the case of KIRP (tumoral tissue) in which a quasi-clique is emerged (see Figure 2). For this reason, we have considered all the CpG islands in our networks as relevant features to compare with the novel feature extraction method proposed in this paper.

We mapped the extracted CpG islands to the genes and we investigated if they are equal to the ones computed by BIGBIOCL. Indeed, when analyzing BRCA we found that seven out of eight genes appear also in the results of BIGBIOCL (AGRN, ISG15, SAMD11, SDF4, SPICE1, TNFRSF18, TNFRSF4). For KIRP two out of six genes appear also in BIGBIOCL (ZNF132, SAMD11), while for THCA no common genes have been identified. For further details the reader may refer to supplementary material S2. We wish to highlight that our method BIGBIOCL extracts many novel genes, which represent additional knowledge with respect to standard correlation analysis.

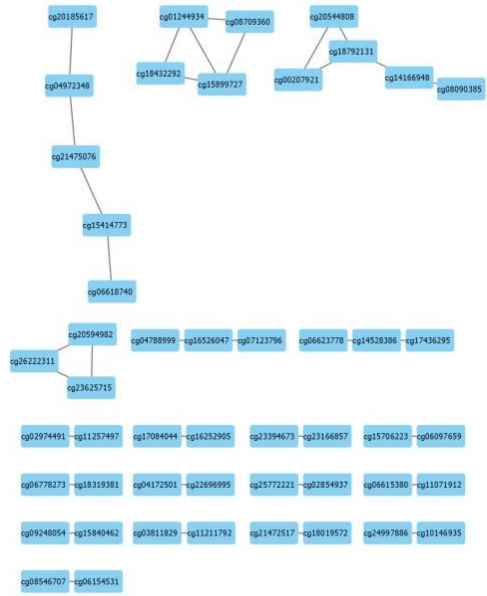

**Figure 1**. Inferred correlation network for the BRCA tumor

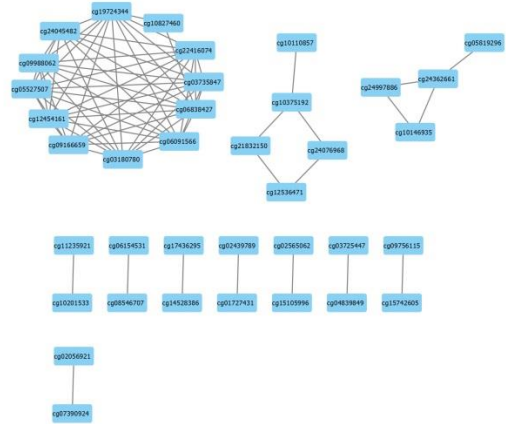

**Figure 2.** Inferred correlation network for the KIRP tumor

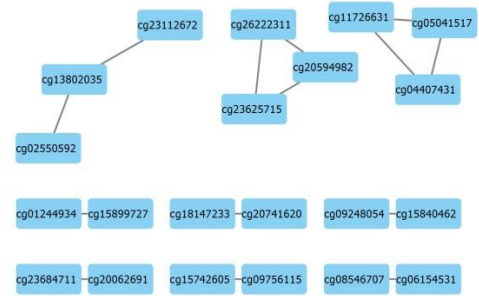

**Figure 3.** Inferred correlation network for the THCA tumor

**Table 17.** Computational time and inferred nodes and edges with the DNA methylation network correlation analysis (Bartlett, *et al.* 2014) implemented in JAVA and executed on the Microsoft Azure Cloud Computing environment using a dual core virtual processor with 14 GB RAM memory and Ubuntu Linux 17.04 operating system

| Disease | Tissue | Experiments | Time | Nodes | Inferred edges |
|---|---|---|---|---|---|
| BRCA | normal | 97 | 3d 20h 20m | 5 | 5 |
| | tumoral | 798 | 18d 12h 57m | 49 | 34 |
| KIRP | normal | 44 | 1d 17h 30m | 0 | 0 |
| | tumoral | 275 | 8d 7h 34m | 37 | 73 |
| THCA | normal | 55 | 2d 3h 44m | 2 | 1 |
| | tumoral | 514 | 11d 1h 4m | 21 | 14 |

## 4　Discussion

Our experiments demonstrate that BIGBIOCL can compute multiple classification models for datasets with hundred thousands of features in few hours. In addition, thanks to the possibility to execute the software on a Hadoop cluster, execution time can be reduced even by 75% compared to Spark local mode. Obviously, the possibility of the software to reach a high level of parallelism allows adding computational nodes to the cluster when the size of the input dataset explodes. The first parameter that can be tuned to improve the parallelism and performance is the number of trees of the Random Forests. This number should be increased only when there is an increment in the size of the input dataset. Increasing the number of trees causes an increase of the training time, which can be contained by adding more computational nodes to the cluster (in fact, trees can be computed in parallel in different nodes).

We compared BIGBIOCL with standard DNA methylation network analysis and other supervised machine learning methods (i.e., SVM and Naïve Bayes) obtaining new knowledge in terms of extracted CpG sites and related genes. In fact, BIGBIOCL represents a novel approach to DNA methylation data classification. BIGBIOCL performs classification using the entire set of features in the input dataset, even when features are hundreds of thousands. This is made possible by the adoption of big data technologies for the computation of the classification model. Other tools work with a smaller set of features (Cestarelli et al., 2016), or reduce the number of features applying regression algorithms (Danielsson et al., 2015).

Datasets used in our experiments were extracted from TGCA and obtained using the HumanMethylation450 platform. This platform provides beta values for more than 485,000 CpG loci. Even if there are more than 28 million of CpG loci in the human genome, data from HumanMethylation450 cover 99% of RefSeq genes, so it is a good starting point to identify drivers for cancer.

In our work, we have provided a methodology and a software tool to analyze HumanMethylation450 data and even bigger datasets. Then, genes extracted from the execution of BIGBIOCL (available at supplementary data S1) can be used by biologists to determine their relevance in a given type of cancer. If we consider that there are around 25 thousand of genes in human DNA, limiting their number allows focusing the attention of the researcher. Analyzing results of experiments on BRCA data, we can find some genes that are well known in literature for their role in breast cancer. For example, mutations of the tumor suppressor gene TP53 and of PIK3CA have been often associated with BRCA (Kim et al., 2017). In addition, both inherited and de novo mutations of BRCA1 and BRCA2 – which mainly cause inactivity of such genes - have been associated to patients with breast cancer (King et al., 2013; Antonucci et al., 2017). A recent study (Tsai et al. 2017) argues that up-regulation of the BDNF signaling pathway can be associated to triple negative breast cancer cells (i.e. cells that test negative for HER2, estrogen receptors, and progesterone receptors). We have obtained BDNF as result of several experiments (IDs 12, 15, 16, and 17). Furthermore, other genes that are considered high-confidence oncogenic candidates (Zheng et al., 2016) have been extracted with BIGBIOCL, as ALDH3A1, CLDN15, SFN, and ENDOD1.

## 5　Conclusion

In conclusion, BIGBIOCL can efficiently manage large datasets, iteratively building equivalent classification models, extracting features (genes in our experiments where features are CpG loci, but the algorithm can potentially be used with other data), and scaling up with the size of the input dataset. Then, results need to be further validated. The algorithm can be improved. It currently builds the classification model on 70% of input data, using 30% of data as test data (the F-measure on test data is used as stopping condition of the iterations). This choice was important during the development and the test of BIGBIOCL. In order to get more precise results and to avoid to loose information, the algorithm could build classification models on 100% of the input data. In addition, as already said, BIGBIOCL can be applied to other type of data, including other NGS experiments and even bigger datasets. Lastly, BIGBIOCL can be used as a component of a pipeline to give sense to raw data, reducing the entropy and focusing the attention on a smaller set of dimensions.

**Acknowledgements**

We wish to thank the Cineca consortium for assigning us supercomputer resources [grant number HP10CTJZAM] to make possible the big data calculations and for providing the PICO infrastructure to carry on experiments with the Hadoop YARN cluster. This work was partially supported by the ERC Advanced Grant GeCo (datadriven Genomic Computing) [grant number 693174], the MoDiag Regione Lazio Project [grant number A0112-2016-13363], and the SysBioNet Italian Roadmap Research Infrastructures. Additionally, we wish to thank Paola Bertolazzi for supporting this work. Finally, we wish to thank Giulia Fiscon and Federica Conte for the patience demonstrated during the local experimentation phase.
*Conflict of Interest:* none declared.